\documentclass[twocolumn,aps]{revtex4}
\usepackage{graphicx}
\def\dbar{$\bar{d}$ }
\def\He3bar{$^3\overline{He}$}
\def\PRL{Phys. Rev. Lett.}
\def\PRC{Phys. Rev. C}
\def\PLB{Phys. Lett. B}
\def\NPA{Nucl. Phys. A}
\def\etal{\emph{et al.}}
\begin{document}
\title{$\bar{d}$ and $^3\overline{He}$ production in $\sqrt{s}_{NN} = 130$ GeV Au+Au collisions
}

\author{
C.~Adler$^{11}$, Z.~Ahammed$^{23}$, C.~Allgower$^{12}$, J.~Amonett$^{14}$,
B.D.~Anderson$^{14}$, M.~Anderson$^5$, G.S.~Averichev$^{9}$, 
J.~Balewski$^{12}$, O.~Barannikova$^{9,23}$, L.S.~Barnby$^{14}$
% ...
\begin{center}(STAR Collaboration)\end{center}
}
% ...
\affiliation{$^5$University of California, Davis, California 95616}
% ...
\affiliation{$^{9}$Laboratory for High Energy (JINR), Dubna, Russia}
% ...
\affiliation{$^{11}$University of Frankfurt, ~Frankfurt, ~Germany}
\affiliation{$^{12}$Indiana University, Bloomington, Indiana 47408}
% ...
\affiliation{$^{14}$Kent State University, Kent, Ohio 44242}
% ...
\affiliation{$^{23}$Purdue University, West Lafayette, Indiana 47907}
% ...

\date{Aug 20, 2001}
\begin{abstract}
The first measurements of light antinucleus production in Au+Au 
collisions at RHIC are reported.  The observed production rates
for \dbar and \He3bar are much larger than in lower energy nucleus-nucleus collisions.  A coalescence model analysis of the yields indicates that there is
little or no increase in the antinucleon freeze-out volume compared 
to collisions at SPS energy.  These analyses also indicate 
that the \He3bar freeze-out volume is smaller than the \dbar freeze-out volume.
\end{abstract}
\maketitle


The Relativistic Heavy-Ion Collider (RHIC) at Brookhaven National Laboratory
has recently begun operation with Au beams at $\sqrt{s}_{NN} = 130$ GeV and extends the available center of mass energy
in nucleus-nucleus collisions by nearly a factor of 8 over CERN SPS collisions
at $\sqrt{s}_{NN} = 17$ GeV.  First measurements
from RHIC indicate an increase of at least 70\% in the charged multiplicity for central collisions compared to previous measurements \cite{PHOBOS}.  Measurements of the antiproton to proton ratio at mid-rapidity \cite{STARpbarp}
indicate that the central collision region is approaching the net-baryon free limit. 
Such a system with large multiplicity and small net-baryon density is
well suited for the production of light antinuclei.  
In this letter, we report the first measurements 
of \dbar and \He3bar production at RHIC.    
 
At RHIC energies, production of antinuclei is possible
via two mechanisms.  The first mechanism is direct production of nucleus-antinucleus pairs in elementary nucleon-nucleon or parton-parton interactions.  The
RHIC center-of-mass energy is well above the threshold for such processes.
Due to their small binding energies, nuclei or antinuclei produced via early direct production are likely to be dissociated in the 
medium before escaping.  

The second, and presumably dominant, 
mechanism for antinucleus production is via
final-state coalescence \cite{Butler,Schwarzschild,Gutbrod}.  In this picture, produced antinucleons merge to
form light antinuclear clusters during the final stages of kinetic freeze-out.
The measured yield of nuclei or antinuclei with nucleon number $A$ and momentum $P$ is
related to the primordial nucleon invariant yield at momentum $p=P/A$ through a coalescence parameter $B_A$,
\begin{equation}    
E \frac{d^3 N_{A}}{d^{3}P} = B_A (E \frac{d^3 N_{N}}{d^{3}p})^A.
\label{eqn:BA}
\end{equation}
Equation \ref{eqn:BA} requires that antineutrons and antiprotons are produced with identical momentum spectra.  

Previous studies of smaller collision systems have 
noted that the measured coalescence parameter $B_A$ can be directly predicted
from the nuclear wave function of the produced (anti)nucleus \cite{Butler}.  When going
to higher energies or larger collision systems, however, the measured 
coalescence 
parameter is lower than that measured in small systems.  This can be understood by noting that once the collision region is larger than the intrinsic
size of the produced (anti)nucleus, (anti)nucleons of equal velocity are not always in close proximity and hence do not always form a bound state \cite{Sato}.  In this sense, the coalescence parameter can be used to infer the space-time geometry of the system \cite{Mekjian}.  Measurements of light nuclei and  antinuclei 
are thus analogous to two-particle Hanbury-Brown Twiss correlations (HBT) in that they measure  
``homogeneity lengths'' of the system at kinetic freeze-out \cite{Scheibl}.  

The measurements were made using the STAR detector \cite{Ackermann}.  The main tracking detector is a cylindrical Time Projection Chamber (TPC), which resides in a solenoidal magnet that was operated with a field
strength of 0.25 T for the data reported here.  The TPC  
tracks and identifies most charged
particles produced in the central pseudo-rapidity region 
($-1.8<\eta<1.8$) with nearly full azimuthal
coverage.  Events are selected on 
the basis of coincidence of spectator neutron signals in two Zero-Degree 
calorimeters located $\pm$18.25 m from the nominal interaction region.  Central events are selected
using a Central Trigger Barrel (CTB) that measures the charged particle
multiplicity with full azimuthal coverage in the pseudo-rapidity 
region $-1<\eta <1$.  This analysis focuses on semi-central 
events, where the centrality corresponds to roughly the most central 18\% of the measured minimum-bias multiplicity distribution.  
The analysis uses $\approx$ 600,000 
events where the interaction vertex is within the range covered by the TPC ($-200<z<200$ cm).  

\begin{figure}[tb]
\begin{center}
\includegraphics[height=5cm]{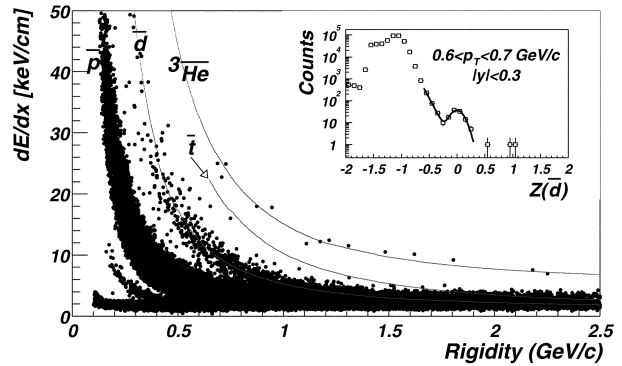}
\end{center}
\caption{Ionization energy loss ($dE/dx$) versus rigidity ($|$momentum/nuclear charge units$|$) for negative tracks.  The $\pi^-$ and $K^-$ bands have been suppressed.  Also plotted are the Bethe-Bloch expectations for $\bar{d}$, $\bar{t}$ and $^3\overline{He}$.  Inserted is a projection of the $Z$ variable (see text) for one transverse momentum bin ($0.6<p_T<0.7$ GeV/c).}
\label{dedx}
\end{figure}

Particle identification is done by measuring the average ionization energy loss ($dE/dx$) for each track.   Studies of the STAR electronics response show
no evidence for saturation below 30 times minimum ionizing.  
For tracks with sufficient transverse momentum to leave the TPC, the path length
exceeds 1.4 m.  For the tracks used in this analysis, the $dE/dx$ resolution is $\approx$11\%. 
For each track,
up to 45 ionization space-point samples are taken along the path through the TPC.  Space-points are found by identifying local maxima of the ADC distribution.
Merged ionization clusters, where multiple tracks
contribute, are identified by looking for multi-peaked structure in the
ADC distribution.   
For the current analysis of relatively rare particles, it is necessary to impose
tight cuts to eliminate background tracks with improperly measured $dE/dx$.
 We require a track to have at least 35 of the 45 possible space-points.  For central events, cluster merging is quite common and can
lead to problems with the particle identification.  To avoid these problems,
we eliminate potentially merged clusters 
from the sample used to calculate the $dE/dx$.   For the final
sample, we require that no more than 30\% of the measured space-points come from potentially merged clusters.  To avoid the Landau tails in the $dE/dx$ spectrum, we use a truncated
mean of the lowest 70\% of the measured $dE/dx$ samples.  Figure \ref{dedx} shows the
measured truncated mean $dE/dx$ versus the magnetic rigidity for the negatively-charged tracks considered
in this analysis.

Figure \ref{dedx} also shows the Bethe-Bloch expectation
for \dbar, $\bar{t}$ and \He3bar.  There is a clear \dbar band below rigidity $\approx$ 1 GeV/c.
This analysis uses only the kinematic region of good \dbar particle identification 
and efficiency ($0.5<p_T<0.8$ GeV/c and rapidity $|y|<0.3$).  
We observe 14 counts clustered around the \He3bar expectation in the
kinematic range $1.0<p_T<5.0$ GeV/c and $|y|<0.8$.  Note
that we plot the rigidity, so the momentum of the \He3bar candidates is twice as large.  No clear $\bar{t}$ band is observed, but
if one assumes that $\bar{t}$ and \He3bar are produced in similar numbers and with similar
momentum distributions we would expect the bulk of the $\bar{t}$  to have 
a higher rigidity where our $dE/dx$ resolution is inadequate for their identification.

To extract the \dbar yield, we construct a 
quantity $Z = log([dE/dx]/I_{\bar{d}}(p))$, 
where $I_{\bar{d}}(p)$ is the expected ionization
for a \dbar of momentum $p$.  For a pure 
sample of $\bar{d}$, this quantity should be well described 
by a Gaussian centered at zero.  In the insert of Fig. \ref{dedx}, we
plot the $Z$ distribution for one transverse momentum bin.  We see a Gaussian 
\dbar signal superimposed
upon a background due to the tail of the $\bar{p}$ distribution.  
We parameterize the $\bar{p}$ background
in the tail region as an exponential, and fit the resulting distribution to a 
Gaussian \dbar
signal + exponential $\bar{p}$ 
tail hypothesis.  In the insert of Fig. \ref{dedx}, we
also show (by the curve) our exponential+Gaussian fit.
In the \dbar kinematic region considered, the signal
to background ratio ranges from 30 in the 
lowest $p_T$ bin to 3 in the highest $p_T$ bin.  
We have performed a similar analysis of the \He3bar 
$Z$ distribution, and estimate the total
background to be less than 0.5 counts.  For extracting yields, we 
assume that our 14 observed \He3bar are background free.   

To evaluate the efficiency, we use GEANT and a TPC response simulator to create
raw pixel level simulated tracks which we then embed into real events.  The embedding
is crucial for this analysis since it allows us to estimate the effects
 of cluster merging
on our efficiency.  No data on \dbar and \He3bar interactions in material exist in the literature, and these antinuclei are 
not incorporated into GEANT. 
Instead we use $d$ and $^3He$ simulations in GEANT
to understand our acceptance and tracking efficiency.  
We then add a correction for
the estimated annihilation in the detector, where we assume that the \dbar
annihilation cross-section is 1.4 times the $\bar{p}$ annihilation cross-section, and that the \He3bar annihilation cross-section is twice the $\bar{p}$
annihilation cross-section.
The $\bar{p}$ annihilation correction was discussed 
in a previous publication \cite{STARpbarp},
and the cross-section scaling relations are taken from Ref. \cite{Hoang}.  
Final calculated efficiencies are in the 
range 0.2-0.5.  
This is much lower than the typical STAR efficiency for charged particle
tracking (0.8-0.9). The
difference is due entirely to the restrictive track cuts used in the 
current analysis
to eliminate backgrounds.      

Systematic errors were estimated by varying the cuts used in the 
analysis.  These variations include changing the number of hits for a valid
track, changing the allowed region of vertex locations, changing the
assumed annihilation cross-sections, and changing the $Z$ range used for the signal+background fit.  We estimate the maximum 
systematic error on the invariant yields to be around 15\%.  We also assume
that the errors on the individual yields are largely correlated.  This causes
the systematic errors to partially cancel when forming coalescence ratios.     

We extract \dbar invariant yields in three transverse momentum bins, 
where each bin has 
$\approx$ 100 entries. 
The extracted yields are listed in Table \ref{InvYield}.
Comparing these yields to lower energies, 
there is a factor of $\approx$ 50 increase
in the $\bar{d}$ production rate in going from
$\sqrt{s}_{NN}=17$ GeV \cite{NA44} to $\sqrt{s}_{NN}=130$ GeV,
and an even more dramatic factor of $\approx$ 60,000 increase in
the $\bar{d}$ production rate relative to 
AGS energy ($\sqrt{s}_{NN}=4.9$ GeV) \cite{E864}.

The mean transverse momentum of the observed \He3bar 
sample is $\approx$2.4 GeV/c.  
We extract an invariant yield per event evaluated at the mean $p_T$ of 
$[8.4\pm2.3(stat.)\pm1.3(sys.)] \times 10^{-7} \mathrm{GeV^{-2}c^3}$. 
We have assumed an exponential transverse mass distribution 
to calculate a cross-section weighted average efficiency in the STAR 
acceptance.  NA52 has reported two \He3bar in 
minimum-bias Pb+Pb collisions at the CERN SPS \cite{NA52}.  Our invariant 
yield is higher, but quantitative comparison cannot be made because of the 
different centralities.

Although only 14 counts were observed, 
our large kinematic coverage for \He3bar allows us to 
estimate the $dN/dy$ and inverse slope $T$.  
To do this, we have calculated the expected yield as a function
of $y$ and $p_T$ using efficiency calculations from embedded data and assuming
an exponential transverse mass distribution.  We minimize the negative 
log-likelihood over the entire STAR acceptance taking 
into account phase-space cells
with no observed counts.  We 
extract \He3bar $dN/dy = [5.1\pm1.7(stat.)\pm0.8(sys.)]\times10^{-5}$ 
and an inverse slope $T=0.70\pm0.25(stat.)$ GeV.  

STAR has measured invariant yields for $\bar{p}$ in a similar centrality range \cite{STARpbar}.  These results
can be combined with the invariant yields presented in this paper to calculate coalescence factors using Equation \ref{eqn:BA}.  In the coalescence picture, only antinucleons produced directly from the source are available to form light
antinuclei. 
Hence, the $\bar{p}$ yields in the coalescence ratio have been corrected for antihyperon feeddown.  We use the 
RQMD model \cite{RQMD} and a detector simulator to evaluate the probability of incorrectly assigning a weak-decay produced
$\bar{p}$ to the primary vertex, and find that about 45$\pm$5(sys.)\% of 
our $\bar{p}$ sample comes from 
antihyperon feeddown.  This fraction is consistent with 
preliminary STAR measurements of the $\bar{\Lambda}/\bar{p}$ ratio.
Table \ref{InvYield} lists the total $\bar{p}$ invariant yields 
along with the estimated correction for antihyperon feeddown.  

\begin{figure}[bht]
\begin{center}
\includegraphics[height=11cm]{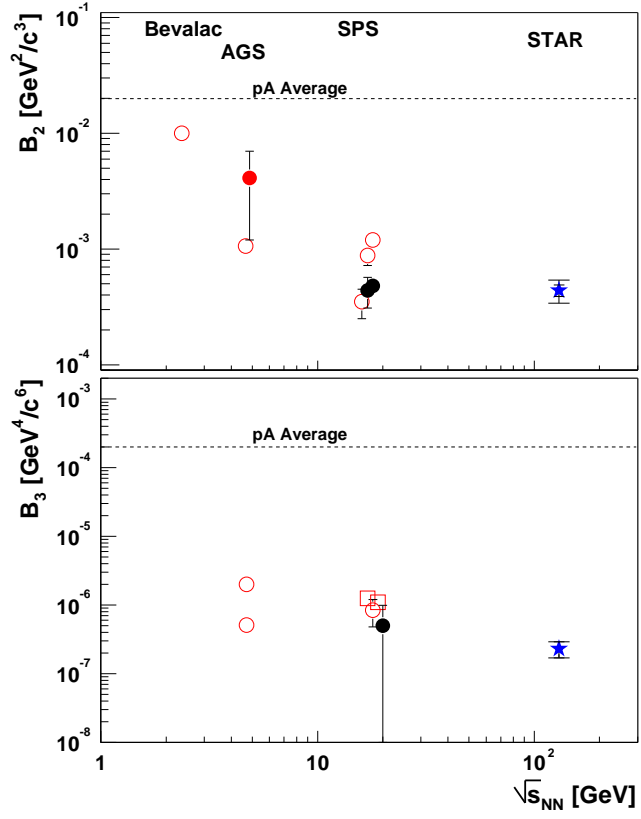}
\end{center}
\caption{Coalescence parameters $B_2$ and $B_3$ excitation functions 
for nuclei (hollow markers) and antinuclei (solid markers) in 
semi-central Au+Au or Pb+Pb collisions \protect{\cite{NA44,E864,EOS,NA52,NA52d,Hansen,NA49,E878}}. The errors on the
STAR data points are statistical (narrow bars) and systematic (wide bars).}
\label{bcomb}
\end{figure}

For the top 18\% most central collisions, we find 
$\langle B_2 \rangle = [4.5\pm0.3(stat.)\pm1.0(sys.)]\times 10^{-4} \; \mathrm{GeV^2/c^3}$ in the $\bar{d}$ kinematic
region $0.5<p_T<0.8$ GeV/c and $|y|<0.3$.  In the top panel of
Fig. \ref{bcomb} we 
compare this result 
to previous measurements at lower energies.  Here we plot the results 
for both $d$ and $\bar{d}$.  In $pA$ collisions,
$B_2$ is essentially independent of the collision energy.  
In central nucleus-nucleus collisions, however, the 
coalescence factor $B_2$ decreases as the collision energy increases from 
Bevalac to AGS to SPS.  
The STAR
result shows that there is no similar decrease in $B_2$ from  
$\sqrt{s}_{NN} = 17$ GeV
to $\sqrt{s}_{NN} = 130$ GeV. Comparing the STAR result
to the average of the two \dbar results at the SPS \cite{NA44,NA52}, we obtain
$B_2(SPS)/B_2(RHIC) = 1.1\pm0.1(stat.)$.

For the top 18\% most central collisions, we find $\langle B_3 \rangle = [2.1 \pm 0.6 (stat.)\pm 0.6(sys.)] \times 10^{-7} \; \mathrm{GeV^4/c^6}$ 
in the \He3bar kinematic region 
$1.0<p_T<5.0$ GeV/c and $|y|<0.8$. 
Once again, we compare to collisions at lower energies in the bottom panel of
Fig. \ref{bcomb}. 
The qualitative trend for $B_3$ is very similar to $B_2$.  For $pA$ 
collisions, the coalescence factor
is independent of energy.  For $AA$ collisions, the coalescence factor 
decreases with increasing collision energy.
The statistics of the \He3bar measurement at the SPS preclude a 
quantitative comparison.  If we compare
to the average of \He3bar and $^3He$ at the SPS \cite{NA52}, we obtain 
$B_3(SPS)/B_3(RHIC) = 3.4\pm1.5(stat.)$.

Several prescriptions have been proposed for relating the coalescence 
parameters to a geometrical source size \cite{Sato,Mekjian,Scheibl,Llope}.  
For these models, the coalescence
parameter scales with the volume as $B_A \propto 1/V^{(A-1)}$ in the limit
of an (anti)nucleon volume much larger than the intrinsic size of the 
produced (anti)nucleus.  Using this simple expression, 
and the measured coalescence parameter ratios, we see that 
$V_{\bar{d}}(RHIC) = (1.1\pm0.1) \; V_{\bar{d}}(SPS)$ and 
$V_{^3\overline{He}}(RHIC) = (1.8\pm0.4) \; V_{^3\overline{He}}(SPS)$.  Both
measurements indicate no large increase 
of the antinucleon
freeze-out volume when going from 
$\sqrt{s}_{NN} = 17$ GeV to $\sqrt{s}_{NN} = 130$ GeV.  
STAR has also measured source sizes 
using $\pi^{-}\pi^{-}$ interferometry \cite{STARhbt}.  
If we construct a quantity proportional to the
volume, $V_{\pi\pi} \propto R_s^2 R_L$, and compare  to the published
SPS data \cite{NA44hbt}, we estimate 
$V_{\pi^-\pi^-}(RHIC) = (1.8 \pm 0.7) \; V_{\pi^-\pi^-}(SPS)$.  
All three available measurements indicate only a slight increase
in volume compared to lower energy collisions.  Caution 
should be exercised, however, when making quantitative comparisons between
the volumes measured via
coalescence and the volumes measured via HBT since it is not clear that
the freeze-out space-time geometry for pions and antinucleons should be the same.   

We can also make quantitive estimates of the freeze-out geometry within
the context of a particular coalescence model and ask 
whether the \dbar and \He3bar sources are the same.  
To do this, we use a simple thermal model \cite{Mekjian}.  This
model assumes that antinucleons and antinuclei are in chemical and thermal
equilibrium within a volume $V$.  From this model, 
we extract $V_{\bar{d}} = 6700 \pm 500$(stat.) fm$^3$, 
and $V_{^3\overline{He}} = 3800 \pm 500$(stat.) fm$^3$.
This discrepancy indicates that the 
thermal model assumptions are not valid
in the production of light antinuclei.  The \He3bar 
freeze-out from a smaller volume and at a presumably earlier time
 compared to \dbar.  This trend of decreasing source size with increasing nucleon number
has been observed before in the production of light nuclei \cite{Mekjian,E864}.
The coalescence picture of light antinucleus production would predict that the
probability for producing an antinucleus with mass $A$ is proportional
to the $A^{th}$ power of the \emph{local} antinucleon density.  
If the 
antinucleon source is not of uniform density, one would expect the different
mass antinuclei to measure different source sizes, and this is indeed what we
observe. 
We have applied other coalescence models with 
different assumptions.  The Sato and Yazaki model \cite{Sato} indicates
a similar trend as the thermal model, with 
$V_{\bar{d}}/V_{^3\overline{He}} = 2.2\pm0.3$, 
while the Scheibl and Heinz model \cite{Scheibl}, 
which can be calculated assuming a Gaussian antinucleon density profile
and explicitly includes the effects of radial flow, 
gives $V_{\bar{d}}/V_{^3\overline{He}} = 0.9\pm0.1$.  In the  Scheibl and
Heinz model, an equivalent effective volume, as indicated by the data,
would imply a larger total volume for \dbar compared to \He3bar.

In summary, we have made the first measurements of the production of light
antinuclei (\dbar and \He3bar) in Au+Au collisions at 
$\sqrt{s}_{NN} = 130$ GeV.  
A large enhancement in production rate is observed compared to 
lower energies.  We have combined the measured yields with
measurements of $\bar{p}$ production to extract coalescence 
parameters $B_2$ and $B_3$. Quantitative comparisons to SPS 
results indicate little or no increase
of the antinucleon freeze-out volume.  
We also find that the \He3bar are produced
from a smaller volume than the \dbar.

We wish to thank the RHIC Operations Group and the RHIC Computing Facility
at Brookhaven National Laboratory, and the National Energy Research 
Scientific Computing Center at Lawrence Berkeley National Laboratory
for their support. This work was supported by the Division of Nuclear 
Physics and the Division of High Energy Physics of the Office of Science of 
the U.S.Department of Energy, the United States National Science Foundation,
the Bundesministerium fuer Bildung und Forschung of Germany,
the Institut National de la Physique Nucleaire et de la Physique 
des Particules of France, the United Kingdom Engineering and Physical 
Sciences Research Council, Fundacao de Amparo a Pesquisa do Estado de Sao 
Paulo, Brazil, and the Russian Ministry of Science and Technology.

\bibliographystyle{unsrt}


\newpage
\begin{table}[tb]
\centering
\begin{tabular}{|c|c|c|c|c|} \hline
 & $\mathrm{p_T (GeV/c)}$ & $\mathrm{E \frac{d^3 N}{d^3p} (GeV^{-2}c^3)}$ & $\mathrm{\bar{p} \; \; E \frac{d^3 N}{d^3(p/A)} (GeV^{-2}c^3)}$ & w.d. correction\\ \hline
      & 0.55 & $[2.47\pm0.26]\times 10^{-3}$ & $4.20\pm0.12$ & $0.56$ \\ \cline{2-5}
\dbar & 0.65 & $[1.87\pm0.19]\times 10^{-3}$ & $4.00\pm0.10$ & $0.53$ \\ \cline{2-5}
      & 0.75 & $[1.93\pm0.20]\times 10^{-3}$ & $3.82\pm0.09$ & $0.52$ \\ \hline
\He3bar & 2.4 & $[8.4\pm2.3] \times 10^{-7}$ & $2.63\pm0.04$ & $0.61$ \\ \hline
\end{tabular}
\caption{Measured invariant yields of antinuclei.  The errors quoted are statistical only.  Systematic errors are estimated to be 15\%.  Also listed are $\bar{p}$ invariant yields at the same velocity, and the weak-decay correction to
the $\bar{p}$ yield estimated from RQMD.}
\label{InvYield} 
\end{table}

\newpage

\end{document}